# Safeguarding the Future of Mobility: Cybersecurity Issues and Solutions for Infrastructure Associated with Electric Vehicle Charging

Md Rakibul Karim Akanda*, Joao Raimundo Queiroz Pires Santana De Oliveira Lima, Amaya Alexandria Holmes, Christina Bonner

Department of Engineering Technology, Savannah State University—Savannah, GA 31404, United States of America. *Corresponding Author email: makan001@ucr.edu

**Abstract**

The development of an ecosystem that balances consumer convenience and security is imperative given the expanding market for electric vehicles (EVs). The vast amount of data that EV charging station management systems (EVCSMSs) give is powered by the Internet of Things (IoT) ecosystem. Intrusion Detection Systems (IDSs), which track network traffic to spot potentially dangerous data exchanges in IT and IoT contexts, are constantly improving in terms of efficacy and accuracy. Intrusion detection is becoming a major topic in academia because of the acceleration of IDS development caused by machine learning and deep learning techniques. The goal of the research presented in this paper is to use a machine-learning-based intrusion detection system with low false-positive rates and high accuracy to safeguard the ecosystem of EV charging stations (EVCS).

## 1. Introduction

Countries are swiftly implementing EV charging stations, which are becoming more and more common in smart cities [1]. IoT is used by these stations to make life easier and provide operators with more control. But because they are online all the time, they are vulnerable to cyberattacks. These attacks can affect the entire EVCS ecosystem, which includes consumers, EVCSs, and the electrical grid. Reliable charging stations are essential, but quick infrastructure development is required to guarantee long-term growth. The EVCS ecosystem's communication protocols regulate communication, yet these elements are susceptible to hackers. International standards for communication between electric cars and charging stations, ISO 15118 and ISO/SAE 21434, specify procedures for safe and effective transactions. ISO/SAE 21434 offers recommendations for cybersecurity in road vehicles, guaranteeing safe systems, networks, and components to guard against cyberattacks and guarantee passenger safety, whereas ISO 15118 concentrates on communication protocols.

EVCS charging stations are being transformed into intelligent systems using the Internet of Things (IoT), allowing user accounting and remote monitoring. This enables consumers to plan their EV charging according to the cost of electricity at night. This poses difficulties, though, because EVCS that are outfitted with IoT technology may be targeted by criminals. Malicious traffic is detected by intrusion detection systems; however, it can be challenging to discern between malicious and regular traffic due to the complex tactics utilized by attackers. Because of the constant connection required to serve customers and the volume of data transferred between nodes, IoT network security is given priority over IT network security [2].

With machine learning techniques like Naive Bayes, Logistic Regression, and Decision Trees being widely utilized to detect network-based threats, artificial intelligence (AI) has been included into intrusion detection systems (IDSs) for Internet of Things (IoT) devices. To increase accuracy and lessen the consequences of feature selection, deep learning techniques are being researched. However, sixteen well-known EVCSMS used by businesses in the US and Europe were found to have serious zero-day vulnerabilities, which resulted in power outages and illegal access. Compared to other elements of the EV ecosystem, EVCSs have gotten less scholarly attention. According to a study in [3], which measured the seriousness of potential vulnerabilities in operational EVCSs, there is a significant risk in live systems used for EV charging that could have an impact on consumers and the power grid. To lessen the impact of such attacks, the study's conclusion included a list of preventative measures that should be applied to susceptible systems. With machine learning techniques like Naive Bayes (NB), Logistic Regression (LR), and Decision Tree (DT) being widely used for detecting network-based threats, artificial intelligence (AI) has been included into

Intrusion Detection Systems (IDS) for Internet of Things (IoT) systems.

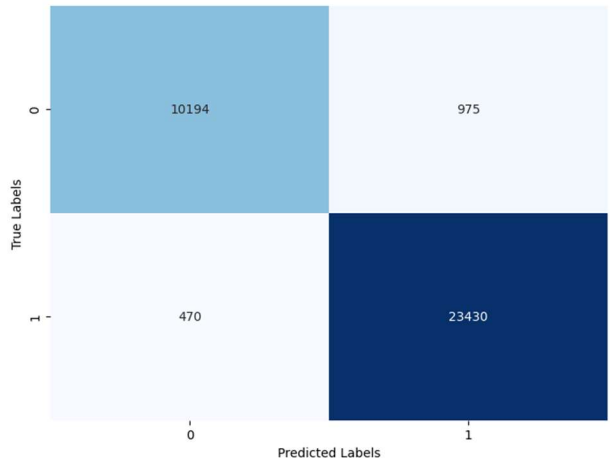

Figure 1. Confusion matrix from Random Forest algorithm

A study by Thakkar et al. [4] highlighted security concerns and hazards related to the application of machine learning and deep learning algorithms in IoT intrusion detection systems. By merging DNN and Long Short-Term Memory (LSTM) neural network learning techniques, they created a deep learning-based intrusion detection system (IDS) to identify denial-of-service (DoS) threats within the EVCS. The Random Forest algorithm was the focus of the authors' comparison of different machine learning techniques using the IoT-23 dataset in [5]. In line with earlier research, they discovered that the Random Forest method had the best accuracy (99.5%).

One useful resource for confirming Internet of Things (IoT) security is the IoT-23 Dataset [6], which was produced using actual commercial IoT network data. This dataset, which was released in 2020 by Avast Software of Prague and the Stratosphere Lab of the Czech Republic, consists of three benign and twenty malicious traffic samples from Internet of Things devices. Malicious traffic from various IoT network assaults is represented by the dataset, which has labels for every potential result.

The usage of machine learning and deep learning to identify abnormalities is growing, but selecting the best algorithm is essential for identifying attacks in real time and stopping harmful communications. Machine learning classifiers are vulnerable to zero-day attacks since they need to be accurate even with small amounts of data. In supervised learning, regularization approaches like L1 and L2 can help solve the overfitting problem [7] and perform better than unsupervised learning algorithms for the same task. Data must be randomized to prevent overfitting, and a reshuffled dataset is run through the Filtered Classifier and Decision Table classifiers to preserve uniformity.

Building and utilizing a basic majority classifier that produces decisions based on several attributes for every instance is the task of decision table classifier rules. Depending on the assignment, the quantity and kinds of qualities can change. The efficacy of the algorithm and calculation time are impacted by the Filtered Classifier's removal of unnecessary data. Prior to more complex techniques like classification and clustering, attribute selection is carried out. The sequential selection of attributes is divided into two phases: ranking, which determines the relative weight of qualities using statistics or information theory, and subset generation, which assesses both determined and possible subsets [8].

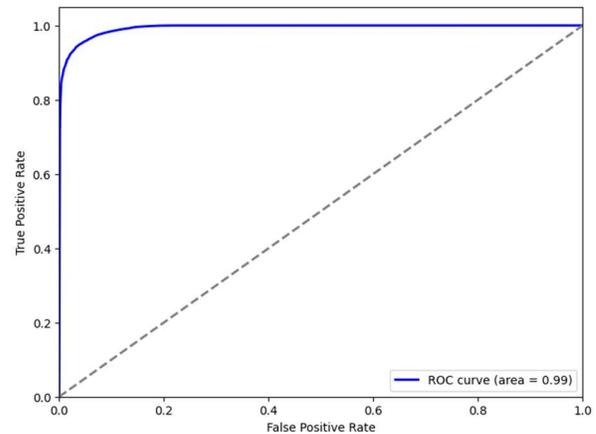

Figure 2. ROC curve from Random Forest algorithm

With China and the US dominating the world in both production and adoption rates, the EV industry has grown remarkably [9]. Government initiatives to encourage clean fuel vehicles and decarbonization initiatives are among the main forces behind the adoption of electric vehicles. Customers are moving toward EVs because of countries like the Netherlands and France announcing prohibitions on the sale of fossil fuel vehicles [10]. Furthermore, range anxiety, a significant obstacle to the adoption of electric vehicles, has been lessened by developments in battery and charging technologies [11]. Long-distance driving has become more practical for electric vehicle users thanks to the widespread availability of high-capacity batteries and rapid charging stations, like those installed by Electrify America in the US and Ionity in Europe [12], [13].

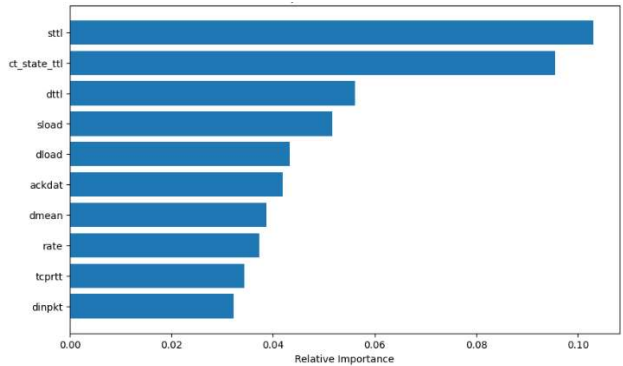

Figure 3. Feature importance from Random Forest algorithm

Alongside these developments, though, there have been worries raised over the cybersecurity of the infrastructure used for charging electric vehicles. Concerns over possible assaults have been raised by recent investigations that found security vulnerabilities in the hardware and applications used for charging electric vehicles. For instance, remote attackers may be able to interfere with the charging of electric vehicles due to flaws in Schneider Electric's EVlink chargers and the ChargePoint Home smartphone software [14], [15]. Strong cybersecurity precautions are necessary because vulnerabilities in web apps for electric vehicle charging systems have also been exploited. These vulnerabilities are being addressed, and efforts are underway to standardize cybersecurity procedures for infrastructure used for charging electric vehicles.

Depending on their energy source, electric vehicles (EVs) are divided into three categories: battery electric vehicles (BEVs), plug-in hybrid electric vehicles (PHEVs), and hybrid electric vehicles (HEVs). BEVs only use an electrochemical battery that is charged by the grid via EV charging stations (EVCSs) in homes or businesses. On the other hand, PHEVs use an electrochemical battery in addition to an internal combustion engine (ICE) that runs on fossil fuels, enabling the two power sources to alternate. These cars frequently have 12- or 24-volt auxiliary power systems that run extra loads and controls. To develop electronics that could be used in electric car applications, such as batteries or infrastructure for charging stations, a variety of technologies and materials have been researched [16–29]. In addition to batteries, some EVs have supercapacitors to speed up energy transfer during regenerative breaking and ignition. Batteries are still the main energy storage component in EVs, nevertheless, because of their lower energy density. Developing effective cybersecurity measures to protect EVs and the larger transportation ecosystem requires an understanding of the EV cyber layers and the dangers they pose.

## 2. Procedure

This paper aims to analyze a cybersecurity dataset [30] to predict whether an attack occurs, using logistic regression and random forest models. The program processes the data, trains the models, evaluates their performance, and visualizes the results through various plots, including confusion matrices, ROC curves, and feature importance charts. It also generates a filtered correlation heatmap to understand the relationships between selected features. Explanations of model are described below.

*Logistic Regression:* This is a statistical model used for binary classification problems. It estimates the probability that a given input point belongs to a certain class. We chose logistic regression because it is simple, interpretable, and effective for binary classification tasks.

*Random Forest:* This ensemble learning method constructs multiple decision trees and merges them with a more accurate and stable prediction. We used random forest because it handles large datasets well, reduces overfitting, and provides insights into feature importance.

These models were chosen to provide a balance between simplicity and complexity, offering both interpretability and robust performance on the classification task. Detailed Overview of the Program is given below.

i. *Loading the Dataset:* The dataset is loaded from a CSV file using pandas.
ii. *Data Preprocessing:* Unnecessary columns are dropped, and features and the target variable are separated.
iii. *One-Hot Encoding:* Categorical features are one-hot encoded to convert them into a numerical format suitable for machine learning models.
iv. *Data Splitting:* The dataset is split into training and testing sets to evaluate the models' performance on unseen data.
v. *Model Training and Evaluation:* Logistic Regression and Random Forest models are

trained, evaluated, and their performance metrics are printed.

vi. *Visualization:*

*Confusion Matrix:* Displays the number of correct and incorrect predictions.

*ROC Curve:* Shows the trade-off between true positive rate and false positive rate.

*Feature Importances:* For Random Forest, the relative importance of each feature is plotted.

vii. *Correlation Heatmap:* Visualizes the correlation between selected features to understand their relationships.

*Confusion Matrix Description:* This confusion matrix (Figure 1) shows the actual versus predicted classifications.

*Interpretation:* Each cell represents the number of predictions for each combination of true and predicted class labels. High values on the diagonal indicate good model performance.

*ROC Curve Description:* A graphical plot (Figure 3) that illustrates the diagnostic ability of a binary classifier.

*Interpretation:* The area under the ROC curve (AUC) represents the degree of separability achieved by the model. Higher AUC values indicate better performance.

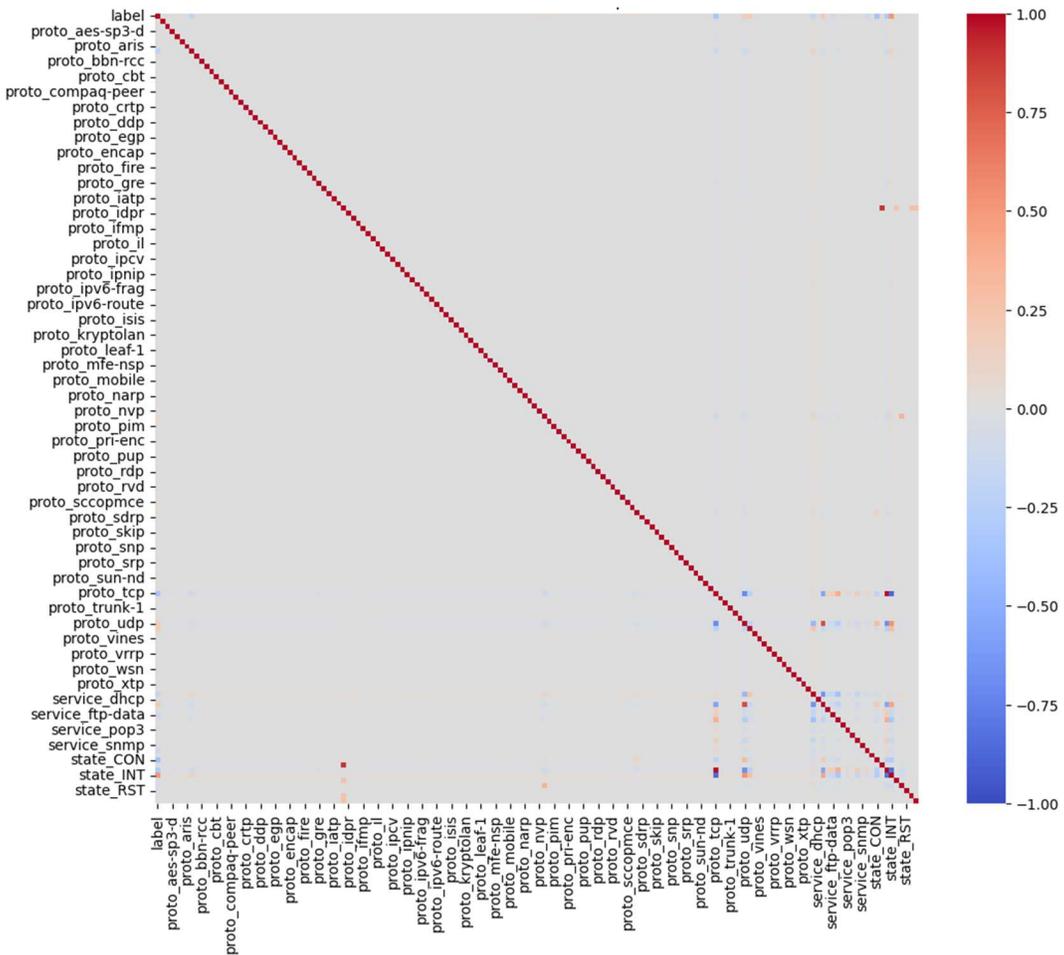

Figure 4. Correlation heatmap

*Feature Importances Description:* This bar plot (Figure 3) shows the top features contributing to the model's predictions.

*Interpretation:* Features with higher importance scores have a greater impact on the model's decisions. This insight helps in understanding which features are most influential.

*Correlation Heatmap Description:* A heatmap (Figure 4) that displays the correlation coefficients between selected features.

*Interpretation:* Values closer to 1 or -1 indicate strong positive or negative correlations, respectively. This helps in identifying redundant features or understanding feature relationships.

**Python Program:**

```python
import pandas as pd
import matplotlib.pyplot as plt
from sklearn.model_selection import train_test_split
from sklearn.metrics import confusion_matrix, classification_report, roc_curve, auc
from sklearn.preprocessing import OneHotEncoder
from sklearn.compose import ColumnTransformer
from sklearn.decomposition import TruncatedSVD
from sklearn.feature_selection import SelectKBest, chi2
from sklearn.linear_model import LogisticRegression
from sklearn.ensemble import RandomForestClassifier
import seaborn as sns

# Step 1: Load the dataset
df = pd.read_csv('/content/cybersecurity_data.csv')

# Step 2: Preprocess the data
# Drop columns that are not needed for prediction
df = df.drop(['id', 'attack_cat'], axis=1)

# Separate features and target variable
X = df.drop(['label'], axis=1)
y = df['label']

# Identify categorical columns to be one-hot encoded
categorical_cols = ['proto', 'service', 'state']

# Apply one-hot encoding to categorical columns
preprocessor = ColumnTransformer(
    transformers=[('cat', OneHotEncoder(), categorical_cols)],
    remainder='passthrough'
)

X_processed = preprocessor.fit_transform(X)

# Split the data into training and testing sets
X_train, X_test, y_train, y_test = train_test_split(X_processed, y, test_size=0.2, random_state=42)

# Define models to evaluate
models = {
    "Logistic Regression": LogisticRegression(max_iter=10000),
    "Random Forest": RandomForestClassifier(n_estimators=100, random_state=42)
}

# Train, evaluate, and visualize results for each model
for model_name, model in models.items():
    print(f"Evaluating {model_name}...")

    # Initialize and train the model
    model.fit(X_train, y_train)

    # Make predictions
    y_pred = model.predict(X_test)
    if hasattr(model, "predict_proba"):
        y_pred_proba = model.predict_proba(X_test)[:, 1]
    else:
        y_pred_proba = model.decision_function(X_test)

    # Evaluate the model
    print(f'Classification Report for {model_name}:")
    print(classification_report(y_test, y_pred))

    # Confusion matrix
    cm = confusion_matrix(y_test, y_pred)
    plt.figure(figsize=(8, 6))
    sns.heatmap(cm, annot=True, fmt='d', cmap='Blues', cbar=False)
    plt.title(f'Confusion Matrix for {model_name}')
    plt.xlabel('Predicted Labels')
    plt.ylabel('True Labels')
    plt.show()

    # ROC Curve
    fpr, tpr, _ = roc_curve(y_test, y_pred_proba)
    roc_auc = auc(fpr, tpr)

    plt.figure(figsize=(8, 6))
```

```
    plt.plot(fpr, tpr, color='blue', lw=2, label=f'ROC curve (area = {roc_auc:.2f})')
    plt.plot([0, 1], [0, 1], color='gray', lw=2, linestyle='--')
    plt.xlim([0.0, 1.0])
    plt.ylim([0.0, 1.05])
    plt.xlabel('False Positive Rate')
    plt.ylabel('True Positive Rate')
    plt.title(f'Receiver Operating Characteristic (ROC) Curve for {model_name}')
    plt.legend(loc='lower right')
    plt.show()

    # Visualize feature importances (only for Random Forest)
    if model_name == "Random Forest":
        importances = model.feature_importances_
        indices = importances.argsort()[::-1][:10]

        # Get feature names
        one_hot_features = preprocessor.transformers_[0][1].get_feature_names_out() if hasattr(preprocessor.transformers_[0][1], 'get_feature_names_out') else []
        num_features = [col for col in X.columns if col not in categorical_cols]
        feature_names = list(one_hot_features) + num_features

        plt.figure(figsize=(10, 6))
        plt.title(f"Feature Importances for {model_name}")
        plt.barh(range(len(indices)), importances[indices], align="center")
        plt.yticks(range(len(indices)), [feature_names[i] for i in indices])
        plt.xlabel("Relative Importance")
        plt.gca().invert_yaxis()
        plt.show()

# Filtered Correlation Heatmap
# Create a smaller dataset for correlation heatmap visualization
df_small = df[['proto', 'service', 'state', 'label']].copy()
df_small = pd.get_dummies(df_small, columns=['proto', 'service', 'state'])

# Compute correlation matrix
corr_matrix = df_small.corr()

# Plot heatmap for the correlation matrix
plt.figure(figsize=(12, 10))
sns.heatmap(corr_matrix, annot=False, cmap='coolwarm', vmin=-1, vmax=1)
plt.title('Correlation Heatmap')
plt.show()
```

### 3. Conclusion

The logistic regression and random forest models provided insightful results on predicting cyber-attacks. While logistic regression offers interpretability and simplicity, random forest gives robust performance and valuable insights into feature importances. Visualizations, including confusion matrices, ROC curves, and feature importance plots, effectively demonstrated model performance and the significance of various features. The correlation heatmap further aided in understanding feature relationships. Overall, the chosen models and visualization techniques provided a comprehensive analysis of the data set highlighting the critical factors in predicting cyber-attacks.

### Acknowledgements

This work was supported as part of the Modeling and Simulation Program (MSP) grant funded by the US Department of Education under Award No. P116S210002 and Improving Access to Cyber Security Education for Underrepresented Minorities funded by the US Department of Education under Award No. P116Z230007.